\documentclass[12pt, elsarticle, a4paper,final]{iopart}
\usepackage{iopams}

\expandafter\let\csname equation*\endcsname\relax
\expandafter\let\csname endequation*\endcsname\relax

\usepackage{amsmath}
\usepackage{amsfonts}
\usepackage{amsmath}
\usepackage{amssymb}
\usepackage{amsthm}
\usepackage[utf8]{inputenc}
\usepackage{graphicx}
\usepackage{bm}
\usepackage{bbm}
\usepackage{cite}
\usepackage{algorithm}
\usepackage{algpseudocode}
\usepackage{mathrsfs}

\usepackage[breaklinks=true,colorlinks=true,linkcolor=blue,urlcolor=blue,citecolor=blue]{hyperref}

\newcommand{\defeq}{\mathrel{\mathop:}=}

\newcommand{\params}{\boldsymbol{\lambda}}

\begin{document}

	\title[Kappa distribution from Gibbs-Sampling algorithm]%
	{Generation of $N$-particle kappa-distributed velocities via superstatistical Gibbs sampling}

\author{Leonardo Herrera-Fuenzalida$^{1}$\footnote[1]{Corresponding author.} and Sergio Davis$^{2,1}$}
\address{$^1$Departamento de Física y Astronomía, Facultad de Ciencias Exactas, Universidad Andrés Bello, Sazié 2212, piso 7, 8370136, Santiago, Chile}
\address{$^2$Research Center in the Intersection in Plasma Physics, Matter and Complexity (P$^2$mc), Comisión Chilena de Energía Nuclear, Casilla 188-D, Santiago, Chile}

\ead{l.herrerafuenzalida@uandresbello.edu}
\ead{sergio.davis@cchen.cl}

\begin{abstract}
The \emph{kappa} distribution is widely used to model heavy-tailed, non-equilibrium velocity distributions in space and laboratory plasmas. Its correct use in particle-in-cell (PIC) simulations requires more than sampling the single-particle marginal: the $N$-particle \emph{kappa} distribution predicts non-trivial correlations between the kinetic energies of particles.
Building on the superstatistical structure, we implement a Gibbs sampler whose stationary distribution is the joint distribution of the velocities and $\beta$, so that the sampled velocities follow the $N$-particle \emph{kappa} distribution and reproduce these correlations by construction.
\end{abstract}

\section{Introduction}
Velocity distributions in collisionless space and laboratory plasmas frequently depart from the Maxwellian equilibrium and are well described instead by the \emph{kappa} distribution~\cite{Livadiotis2017,Lazar2021}.
For a single particle with mass $m$ and velocity $\bm v$, the kappa distribution is commonly presented in the form
\begin{equation}
	P(\bm v|\kappa, v_{th}) = \frac{1}{\eta_{\kappa} (v_{th})} \left[1+\frac{1}{\kappa-\frac{3}{2}}\frac{\bm v^2}{v^2_{th}} \right]^{-(\kappa+1)}
	\label{kappa_distribution_k}
\end{equation}
where $\kappa > 3/2$ is the shape parameter, $v_{th} := \sqrt{\frac{2k_BT}{m}}$ is a scale parameter related to the effective temperature and $\eta_{\kappa} (v_{th})$ is a normalization constant given by 
 \begin{equation}
 	\eta_{\kappa}(v_{th}) := \left[ \pi v ^2_{th}\left(\kappa- \frac{3}{2}\right)\right]^{\frac{3}{2}} \frac{\Gamma(\kappa-\frac{1}{2})} {\Gamma(\kappa+1)}.
 \end{equation}  
%
%
To initialize simulations of non-equilibrium plasmas with the particle-in-cell (PIC) method, samples of velocities that follow the \emph{kappa} distribution for $N$ particles are required. The simplest approach, drawing each particle's velocity independently from the single-particle \emph{kappa} distribution, reproduces the correct single-particle statistics but not the joint $N$-particle \emph{kappa} distribution, which does not factorize into its single-particle marginals~\cite{Livadiotis2011,Livadiotis2017ch1}
\begin{equation}
	P(\bm V) \neq \prod_{i=1} ^N P(\bm v_i),
\end{equation}
where $\bm V$ collects the velocities $\bm v_i$ of the $N$ particles.
%
%
Two formalisms are commonly invoked to explain the origin of the kappa distribution. In nonextensive statistical mechanics, the \emph{kappa} distribution arises from the maximization of the Tsallis entropy~\cite{Tsallis1988} under \emph{ad hoc} constraints on the moments of the velocities. In superstatistics~\cite{Beck2003,Beck2004,Davis2026}, the inverse temperature $\beta\defeq 1/(k_BT)$ is not a fixed quantity; instead, it is represented by a probability distribution $P(\beta|\lambda)$, so that the \emph{kappa} distribution is obtained as a mixture of Maxwellians weighted by $P(\beta|\lambda)$. This explicit probabilistic structure allows moments and correlations to be computed directly, and makes the superstatistical formulation suited to the construction of sampling algorithms.

In this work we use this formalism to construct a Gibbs sampler~\cite{Gelfand2000} for the $N$-particle \emph{kappa} distribution, which alternates between the conditional distribution of the velocities given $\beta$ (a product of independent Maxwellians) and the conditional distribution of $\beta$ given the velocities (a gamma distribution~\cite{Davis2026}). Since both conditional distributions can be sampled exactly, every step is a direct draw, with no proposal and no rejection.

We show that the sampler reproduces the \emph{kappa} joint distribution of velocities for $N$ particles, the distribution of $\beta$, the distribution of the kinetic energy of a single particle, the joint distribution of the kinetic energies of two particles, and the Pearson correlation and mutual information between them, in agreement with the analytical expressions. 
\section{Kappa distributions in the framework of superstatistics}
\label{sec:theory}

\subsection{Superstatistics}
\label{sec:superstatistic}

Let $\bm{V}\defeq(\bm v_1,\bm v_2,\ldots,\bm v_N)$ be the $3N$-dimensional vector containing the velocity degrees of freedom of $N$ particles, each with three velocity components. At fixed inverse temperature $\beta$, the joint distribution of $\bm V$ is the product of $N$ independent Maxwell-Boltzmann distributions,
\begin{equation}
P(\bm V|\beta) = \prod_{i=1}^N \left(\frac{m_i\beta}{2\pi}\right)^{\frac{3}{2}}\exp\left(-\frac{\beta m_i}{2}\bm{v}_i^2 \right) = \frac{\exp\big(-\beta K(\bm V)\big)}{Z_N(\beta)},
\label{eq:N-maxwell}
\end{equation}
where $K(\bm V) \defeq \sum_{i=1}^N \frac{m_i \bm v^2_i}{2}$ is the total kinetic energy and $Z_N(\beta) \defeq \prod_{i=1}^N (2\pi/m_i\beta)^{3/2} = (2\pi/\tilde m\beta)^{3N/2}$ is the partition function, with $\tilde m$ the geometric mean of the particle masses.

In superstatistics the observed distribution is a superposition of these Maxwellians poderated by the marginal distribution of $\beta$. For a single particle its distribution is
\begin{equation}
P(\bm v|\params) = \int_0^\infty d\beta\, f(\beta;\params)\exp\left(-\frac{\beta m\bm v^2}{2}\right),
\label{eq:superstat}
\end{equation}
with $f(\beta;\params)\defeq P(\beta|\params)/Z(\beta)$ and $\params$ the parameters of the distribution of $\beta$. For $N$ particles, the expression is
\begin{equation}
	P(\bm V| \params) = \int_0^{\infty} d\beta\, \frac{P(\beta|\params)}{Z_N(\beta)} \prod_{i=1}^N \exp \left(-\frac{\beta m_i \bm v_i^2}{2}\right),
	\label{eq:N_superstat}
\end{equation}

\subsection{Kappa distribution in the $(u,\beta_S)$ parametrization}
\label{sec:kappa-ubetaS}

The \emph{kappa} distribution arises when $\beta$ itself follows a $\mathrm{Gamma}(\beta|\params)$ distribution. The parametrization introduced in Ref.~\cite{Davis2023e} uses $1/u$ as the shape and $u\beta_S$ as the scale; throughout this work, $\mathrm{Gamma}(a,b)$ denotes a gamma distribution with shape $a$ and scale $b$. Both parameters are given directly by the moments of $\beta$, since $u \defeq \langle (\delta \beta)^2 \rangle / \langle \beta \rangle ^2$ is the relative variance of $\beta$ and $\beta_S\defeq \langle\beta\rangle$ its mean,
\begin{equation}
P(\beta|u,\beta_S) = \frac{1}{u\beta_S\,\Gamma(1/u)}\exp\left(-\frac{\beta}{u\beta_S}\right)\left(\frac{\beta}{u\beta_S}\right)^{\frac1u-1}, \qquad u>0.
\label{eq:pbeta}
\end{equation}
After marginalizing $P(\bm V|\params)=\int_0^\infty d\beta\,P(\beta|\params)P(\bm V|\beta)$, we obtain the \emph{kappa} distribution in terms of the parameters of the distribution of $\beta$, given by
\begin{equation}
	P(\bm v \mid u, \beta_S) = \left( \frac{m u \beta_S}{2 \pi} \right)^{\tfrac{3}{2}} \frac{\Gamma \left(\tfrac{1}{u}+\tfrac{3}{2}\right)}{\Gamma\left(\frac{1}{u} \right)}  \left[1 + u \beta_S \frac{m \bm v^2}{2} \right]^{-\left( \tfrac{1}{u} + \tfrac{3}{2} \right)},
	\label{eq:kappa_1}
\end{equation}
and for $N$ particles, the expression is
\begin{equation}
	\label{eq:kappa_N-distribution}
	P(\bm V \mid u, \beta_S) = \left(\frac{\tilde m u\beta_S}{2\pi}\right)^{\tfrac{3N}{2}} \frac{\Gamma(\tfrac{1}{u} + \tfrac{3N}{2} )}{\Gamma\left(\tfrac{1}{u}\right)}\left[1 + u\beta_S K(\bm V)\right]^{-\left(\tfrac{1}{u} + \tfrac{3N}{2}\right)}
\end{equation}
with $K(\bm V)$ the total kinetic energy defined in Eq.~\eqref{eq:N-maxwell}. The conventional spectral index of Eq.~\eqref{kappa_distribution_k} is given by $\kappa=1/u+1/2$. The mean kinetic energy is finite for $u<1$ ($\kappa>3/2$), and its variance for $u<1/2$. In the limit $u\to0$, Eq.~\eqref{eq:kappa_N-distribution} reduces to the $N$-particle Maxwellian in Eq.~\eqref{eq:N-maxwell}.
%
%
\subsection{Correlations in the superstatistical \emph{kappa} distribution}
\label{sec:correlations-theory}

In the Maxwellian regime, $\beta$ takes a single value and the joint distribution factorizes, $P(\bm V\mid \beta) = P(\bm v_1|\beta) \cdots P(\bm v_N|\beta)$. In the superstatistical construction, instead, the uncertainty in $\beta$ is described by a single distribution $P(\beta|u,\beta_S)$ for the whole set of particles. Since this distribution is integrated over once for all the particles, and not separately for each of them, the joint distribution in Eq.~\eqref{eq:kappa_N-distribution} depends on the velocities only through the total kinetic energy $K(\bm V)$, and it does not factorize into its single-particle marginals,
\begin{equation}
	P(\bm V \mid u, \beta_S) \neq \prod_{i=1}^{N} P(\bm v_i \mid u,\beta_S), \qquad u>0.
	\label{eq:nonfactor}
\end{equation}
To see this, note first that integrating Eq.~\eqref{eq:N_superstat} over $N-1$ particles, only integrates the Maxwellian factor of particle; the marginal of any single particle is therefore the kappa distribution of Eq.~\eqref{eq:superstat}. Evaluating both at the origin for $N=2$, gives
\begin{equation}
	\frac{P(\bm v_1=\bm 0,\bm v_2=\bm 0\mid \params)}{P(\bm v_1=\bm 0\mid \params)\,P(\bm v_2=\bm 0\mid \params)}=\frac{\langle\beta^{3}\rangle}{\langle\beta^{3/2}\rangle^{2}} = 1+\frac{\mathrm{Var}(\beta^{3/2})}{\langle\beta^{3/2}\rangle^2}>1,
	\label{eq:ratio-origin}
\end{equation}
since $\langle\beta^{3}\rangle-\langle\beta^{3/2}\rangle^{2}$ is the variance of $\beta^{3/2}$, which vanishes only in the Maxwellian limit $u\to0$.

As a consequence, the $N$-particle kappa distribution implies a positive correlation between the kinetic energies of any two particles, even though no interaction between them is involved~\cite{Davis2023e,Davis2026}.

\subsection{Observables of the $N$-particle \emph{kappa} distribution}
\label{sec:observables}

Single-particle observables cannot distinguish the $N$-particle \emph{kappa} distribution from independent particles, since both share the marginal in Eq.~\eqref{eq:kappa_1}, and differ only in the joint structure discussed in Section~\ref{sec:correlations-theory}. We therefore derive closed-form expressions for the kinetic energies at two levels: single-particle quantities, which any correct generator must reproduce, and two-particle quantities, which are sensitive to this joint structure. These expressions are the reference for the validation in Section~\ref{sec:validation}.

\subsubsection{Kinetic energy of a single particle}
\label{sec:obs-energy}

Setting $N=1$ in Eq.~\eqref{eq:kappa_N-distribution}, we obtain the distribution of the kinetic energy of a single particle, $k\defeq\frac{m}{2}\bm v^2$ (Eq.~66 of Ref.~\cite{Davis2026}). After the change of variable $t\defeq u\beta_S k$, we recognize it as a Beta-prime distribution, $t\sim\mathrm{BetaPrime}(3/2,\,1/u)$. In particular, the mean kinetic energy per particle is
\begin{equation}
\langle k\rangle = \frac{3}{2\beta_S(1-u)}, \qquad u<1.
\label{eq:mean-k-d}
\end{equation}

\subsubsection{Joint distribution of the kinetic energies of two particles}
\label{sec:obs-joint}

The joint distribution of the kinetic energies $k_1$ and $k_2$ of two particles carries the non-factorization of Eq.~\eqref{eq:nonfactor} over to observable quantities. Changing variables from velocities to kinetic energies in Eq.~\eqref{eq:kappa_N-distribution} with $N=2$, we obtain (Eq.~68 of Ref.~\cite{Davis2026})
\begin{equation}
P(k_1,k_2\mid u,\beta_S) = \frac{4}{\pi}\,\beta_S^3(1+u)(1+2u)\sqrt{k_1k_2}\,\big[1+u\beta_S(k_1+k_2)\big]^{-(1/u+3)}.
\label{eq:joint-3}
\end{equation}
By maximizing Eq.~\eqref{eq:joint-3}, we obtain the mode of the joint distribution, located at $k_1=k_2=k^*$ with
\begin{equation}
k^* = \frac{1}{2\beta_S(1+2u)}.
\label{eq:jointmode}
\end{equation}
\subsubsection{Pearson correlation}
\label{sec:obs-pearson}

The Pearson correlation summarizes the dependence between $k_1$ and $k_2$ in a single number. Conditioning on a large $k_1$ shifts the distribution of $\beta$ toward smaller values, which raises the expected value of $k_2$. Integrating Eq.~\eqref{eq:joint-3}, the mixed moments of the kinetic energies are
\begin{equation}
\langle k_1^a k_2^b\rangle_{u,\beta_S}=(u\beta_S)^{-(a+b)}\,\frac{\Gamma(\tfrac32+a)\,\Gamma(\tfrac32+b)\,\Gamma(\tfrac1u-a-b)}{\Gamma(\tfrac32)^2\,\Gamma(\tfrac1u)},
\label{eq:mixed-moments}
\end{equation}
which are finite only for $a+b<1/u$. With $a+b\le2$ they give the Pearson correlation coefficient,
\begin{equation}
R(u) \defeq \frac{\langle\delta k_1\,\delta k_2\rangle_{u,\beta_S}}{\langle(\delta k)^2\rangle_{u,\beta_S}} = \frac{3u}{u+2}, \qquad u<\tfrac12,
\label{eq:pearson}
\end{equation}
in agreement with Eq.~63 of Ref.~\cite{Davis2026} and with Ref.~\cite{Davis2023e}. $R$ depends only on $u$, and not on $\beta_S$ or on the mass of the particles.

\subsubsection{Mutual information}
\label{sec:obs-mutinf}

The mutual information measures the total dependence between $k_1$ and $k_2$, linear and nonlinear. Replacing the joint distribution of Eq.~\eqref{eq:joint-3} and the single-particle marginals in its definition, Davis \textit{et al.}~\cite{Davis2026} obtained the closed form (their Eqs.~64-70)
\begin{equation}
I_{k_1,k_2}(u) \defeq \left\langle\ln\frac{P(k_1,k_2|u,\beta_S)}{P(k_1|u,\beta_S)P(k_2|u,\beta_S)}\right\rangle_{u,\beta_S} = 2F\!\left(\tfrac1u+\tfrac32\right)-F\!\left(\tfrac1u\right)-F\!\left(\tfrac1u+3\right),
\label{eq:mutinf}
\end{equation}
where $F(z)\defeq z\psi(z)-\ln\Gamma(z)$ and $\psi$ is the digamma function. Both $R(u)$ and $I_{k_1,k_2}(u)$ vanish for $u\to0$ and grow with $u$.
\subsection{Conditional distribution of $\beta$}
\label{sec:beta-conditional}
Within the probabilistic structure of superstatistics, the temperature is not an observable~\cite{Davis2018}. However, given the vector of velocities, our knowledge of $\beta$ is described by the conditional distribution $P(\beta | \bm V, \params)$, obtained from Bayes' rule,
\begin{equation}
	P(\beta \mid \bm{V}, u,\beta_S) = \frac{P(\bm{V}, \beta \mid u,\beta_S)}{P(\bm{V} \mid u,\beta_S)} \propto P(\beta | \bm V, u, \beta_S).
\end{equation}
In the superstatistical structure, the joint distribution is readily obtained as $P(\bm V|\beta)P(\beta|u,\beta_S)$, and its expression is
\begin{equation}
	\label{join_kappa_distribution}
	P(\bm V, \beta | u, \beta_S) = \left( \frac{\tilde m \beta}{2 \pi}\right)^{\frac{3N}{2}} \frac{1}{u \beta_S \Gamma \left( \tfrac{1}{u}\right)} \exp \left(-\frac{\beta}{u \beta_S}\left[ 1 + u \beta_S K(\bm V) \right] \right) \left( \frac{\beta}{u \beta_S}\right)^{\frac{1}{u}-1},
\end{equation}
and, as a function of $\beta$ with $\bm V$, $u$ and $\beta_S$ fixed, this expression yields the conditional distribution
\begin{equation}
	\label{beta_conditional_distribution}
	P(\beta| \bm V, u, \beta_S) = \frac{1}{\beta_0 \Gamma(\frac{1}{u} + \frac{3N}{2})} \exp \left( -\frac{\beta}{\beta_0} \right) \left( \frac{\beta}{\beta_0}\right)^{\frac{1}{u} + \frac{3N}{2} - 1}
\end{equation}
where $\beta_0 = \frac{u \beta_S}{1 + u \beta_S K(\bm V)}$ \cite{Davis2026}. This is a gamma distribution, $\beta \sim \mathrm{Gamma}(3N/2 + 1/u,\, \beta_0)$, with shape $3N/2+1/u$ and scale $\beta_0$.
\section{The Gibbs sampler for kappa distribution}
\label{sec:gibbs}

\subsection{The general algorithm}

Gibbs sampling~\cite{Gelfand2000,Casella1992} is a Markov chain Monte Carlo method for sampling a joint distribution $P(x_1,\ldots,x_k|\params)$ that cannot be sampled directly, but whose \emph{full conditional distributions} $P(x_j|x_{-j},\params)$, with $x_{-j}\defeq(x_1,\ldots,x_{j-1},x_{j+1},\ldots,x_k)$, are easy to sample. Starting from an arbitrary state $x^{(0)}$, each iteration updates every component $x_j$ (a scalar or a block of variables) in turn, using the most recent values of the others,
\begin{equation}
	x_j^{(t+1)} \sim P\big(x_j \,\big|\, x_1^{(t+1)},\ldots,x_{j-1}^{(t+1)},x_{j+1}^{(t)},\ldots,x_k^{(t)}\big), \qquad j=1,\ldots,k.
	\label{eq:gibbs-sweep}
\end{equation}
Each update leaves the joint distribution invariant~\cite{Geman1984,Gelfand1990} and, under mild conditions, the chain converges in distribution to $P(x_1,\ldots,x_k|\params)$~\cite{Tierney1994,Roberts1994,robert2004monte}, so expectation values are estimated by averaging over the chain after a warm-up segment. Every draw is accepted, so no rejection step or proposal tuning is required. Only the $k$ full conditionals are needed, even when $P(x_1,\ldots,x_k|\params)$ is known only up to a normalization constant~\cite{Gelfand2000}; the procedure is summarized in Algorithm~\ref{alg:gibbs-general}.

\begin{algorithm}[H]
\caption{Gibbs sampling for $k$ random variables $x_1,\ldots,x_k$.}
\label{alg:gibbs-general}
\begin{algorithmic}[1]
\State Begin with a seed $x^{(0)}=(x_1^{(0)},\ldots,x_k^{(0)})$, initialize $t=0$
\For{$j=1,\ldots,k$}
	\State Generate $x_j^{(t+1)}$ from $P\big(x_j \,\big|\, x_1^{(t+1)},\ldots,x_{j-1}^{(t+1)},x_{j+1}^{(t)},\ldots,x_k^{(t)},\params\big)$
\EndFor
\State Save the sample $x^{(t+1)}=(x_1^{(t+1)},\ldots,x_k^{(t+1)})$
\State Make $t=t+1$ and go back to step 2
\end{algorithmic}
\end{algorithm}

\subsection{Gibbs sampling for the superstatistical \emph{kappa} distribution}
\label{sec:algorithm}
\label{sec:conditionals}

Starting from the joint distribution $P(\bm V, \beta | u, \beta_S)$ of Eq.~\eqref{join_kappa_distribution}, whose marginal in $\bm V$ is the \emph{kappa} distribution, the Gibbs sampler alternates between the two full conditional distributions, $P(\bm V |\beta^{(t)})$ in Eq.~\eqref{eq:N-maxwell} and $P(\beta| \bm V^{(t)}, u, \beta_S)$ in Eq.~\eqref{beta_conditional_distribution}. The chain converges to the joint distribution, so that the sampled velocities follow the $N$-particle \emph{kappa} distribution.

\begin{algorithm}[H]
\caption{Gibbs sampling for $N$ kappa particles.}
\label{alg:gibbs-kappa}
\begin{algorithmic}[1]
\State Initialize the value of inverse temperature: $\beta=\beta_S$
\State For each $i=1,2,\ldots,N$:
\Statex \hspace*{3em}Generate the new velocity $\bm v_i$ from the Maxwellian $P(\bm v_i|\beta)$
\Statex \hspace*{3em}Save $\bm v_i$
\State Calculate $\beta_0$ using the current kinetic energy $K(\bm V)=\sum_{i=1}^N \frac{m_i\bm v_i^2}{2}$
\State Update $\beta$ from the distribution $P(\beta|\bm V,u,\beta_S)$
\State Go back to step 2
\end{algorithmic}
\end{algorithm}
\section{Results}
\label{sec:validation}

\subsection{Recovery of the kappa velocity distribution}
\label{sec:res-velocities}

\begin{figure}[H]
\begin{center}
\includegraphics[width=0.6\textwidth]{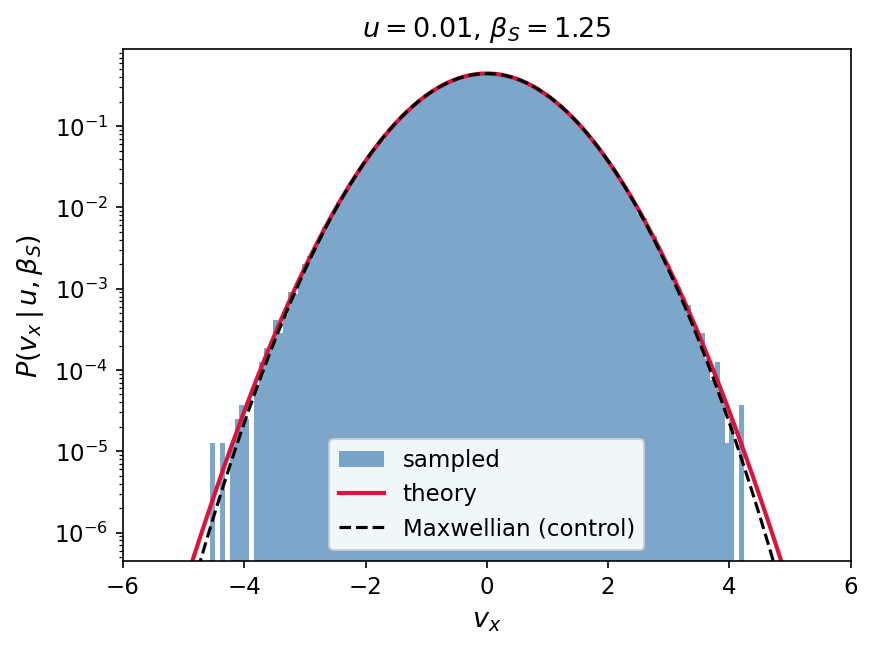}
\end{center}
\caption{Distribution of $v_x$ for $u=0.01$ and $\beta_S=1.25$.}
\label{fig:kappa-velocities-0p01}
\end{figure}

\begin{figure}[H]
\begin{center}
\includegraphics[width=0.6\textwidth]{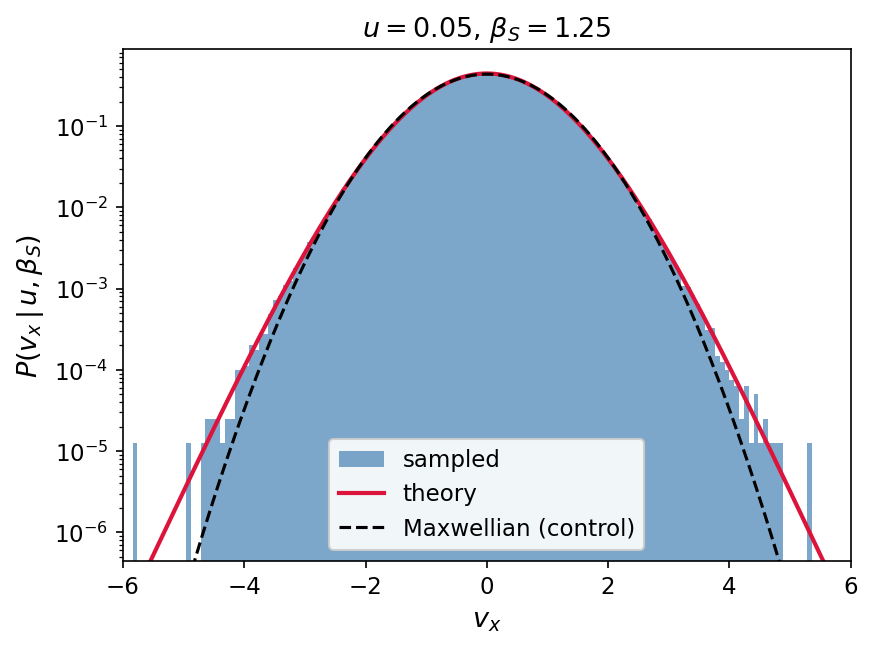}
\end{center}
\caption{Distribution of $v_x$ for $u=0.05$ and $\beta_S=1.25$.}
\label{fig:kappa-velocities-0p05}
\end{figure}

\begin{figure}[H]
\begin{center}
\includegraphics[width=0.6\textwidth]{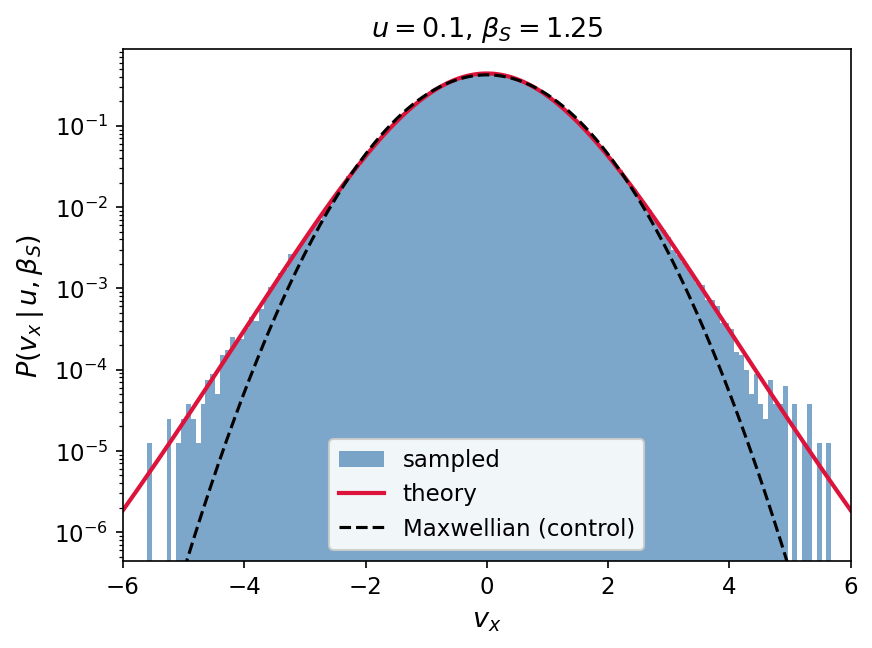}
\end{center}
\caption{Distribution of $v_x$ for $u=0.1$ and $\beta_S=1.25$.}
\label{fig:kappa-velocities-0p1}
\end{figure}

\begin{figure}[H]
\begin{center}
\includegraphics[width=0.6\textwidth]{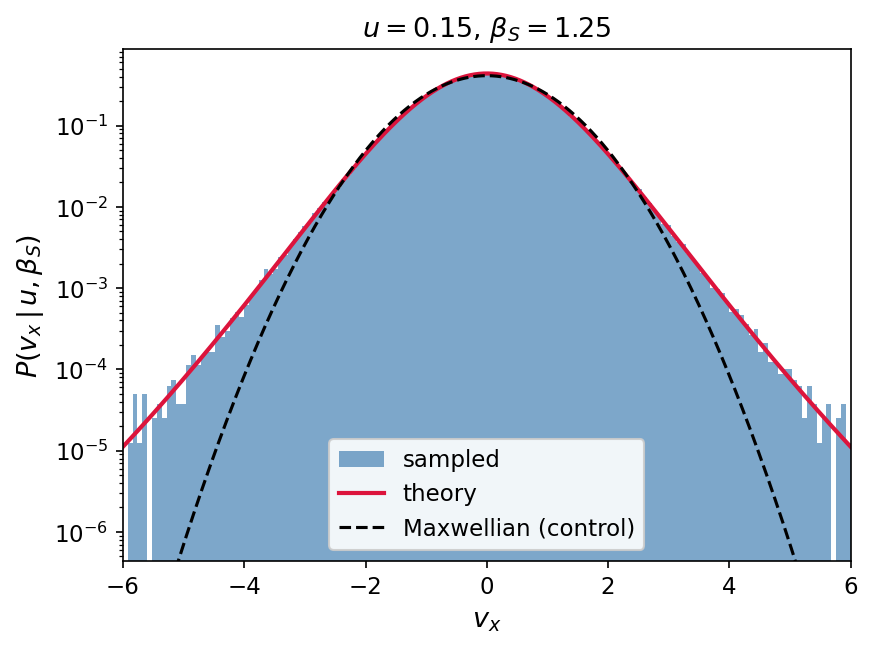}
\end{center}
\caption{Distribution of $v_x$ for $u=0.15$ and $\beta_S=1.25$.}
\label{fig:kappa-velocities-0p15}
\end{figure}

\begin{figure}[H]
\begin{center}
\includegraphics[width=0.6\textwidth]{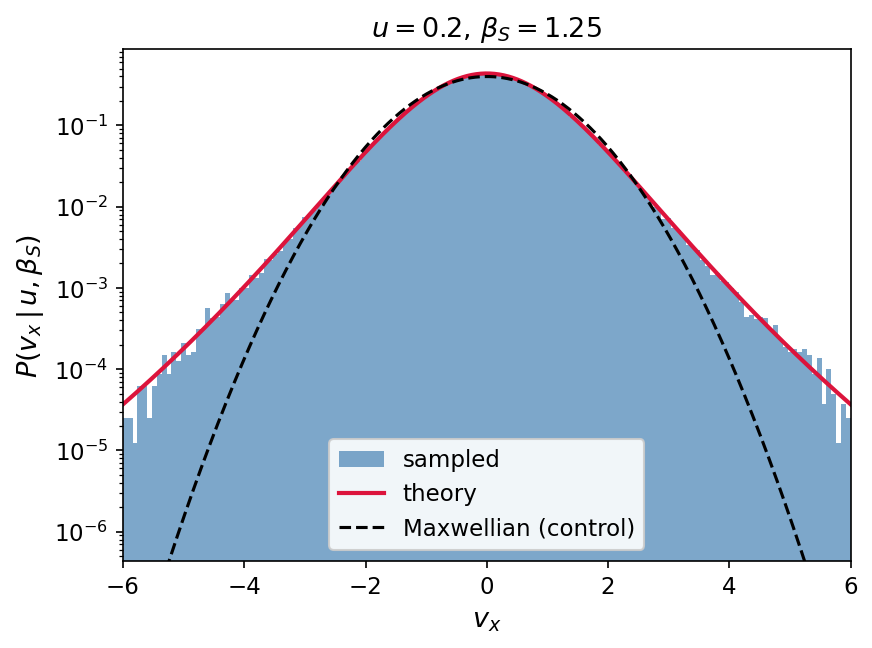}
\end{center}
\caption{Distribution of $v_x$ for $u=0.2$ and $\beta_S=1.25$.}
\label{fig:kappa-velocities-0p2}
\end{figure}

\begin{figure}[H]
\begin{center}
\includegraphics[width=0.6\textwidth]{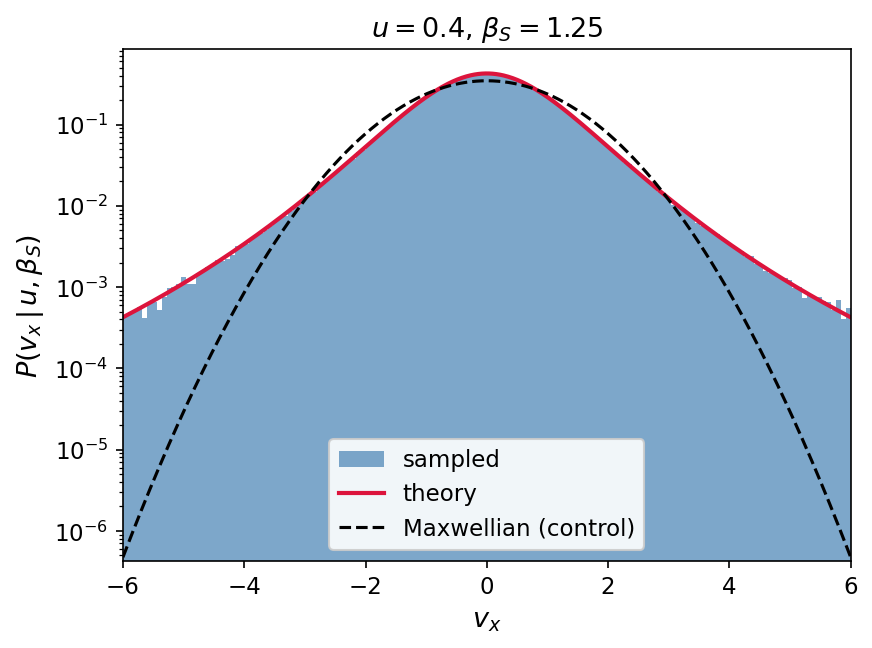}
\end{center}
\caption{Distribution of $v_x$ for $u=0.4$ and $\beta_S=1.25$.}
\label{fig:kappa-velocities-0p4}
\end{figure}

To verify that the sampler recovers the \emph{kappa} distribution of velocities, we compare the histogram of the component $v_x$ of the Gibbs-sampled three-dimensional velocities against its exact \emph{kappa} distribution. Figures~\ref{fig:kappa-velocities-0p01}--\ref{fig:kappa-velocities-0p4} show this comparison for $u\in\{0.01,0.05,0.1,0.15,0.2,0.4\}$, with $\beta_S=1.25$ in all cases. The sampled velocities follow the exact distribution throughout the explored range of $u$, including the tails, which separate increasingly from the Maxwellian control as $u$ grows.

\subsection{Distribution of the inverse temperature $\beta$}
\label{sec:res-beta}

\begin{figure}[H]
\begin{center}
\includegraphics[width=0.6\textwidth]{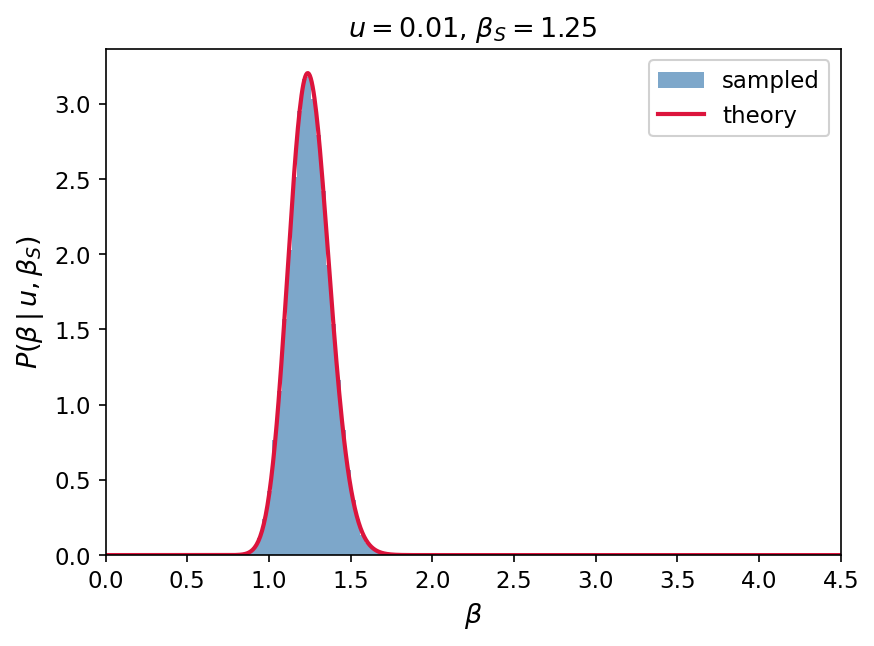}
\end{center}
\caption{Distribution of $\beta$ for $u=0.01$ and $\beta_S=1.25$.}
\label{fig:beta-histo-0p01}
\end{figure}

\begin{figure}[H]
\begin{center}
\includegraphics[width=0.6\textwidth]{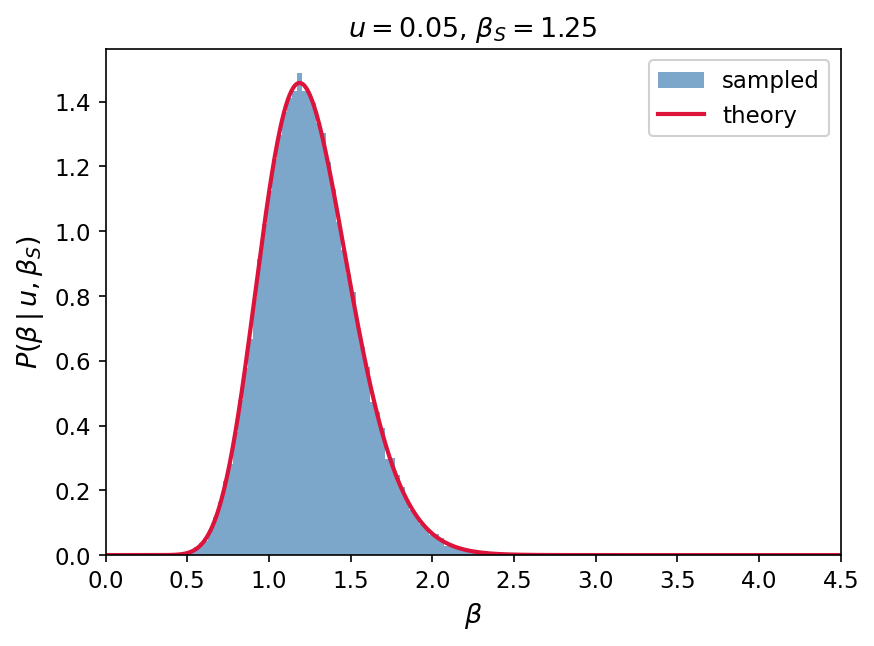}
\end{center}
\caption{Distribution of $\beta$ for $u=0.05$ and $\beta_S=1.25$.}
\label{fig:beta-histo-0p05}
\end{figure}

\begin{figure}[H]
\begin{center}
\includegraphics[width=0.6\textwidth]{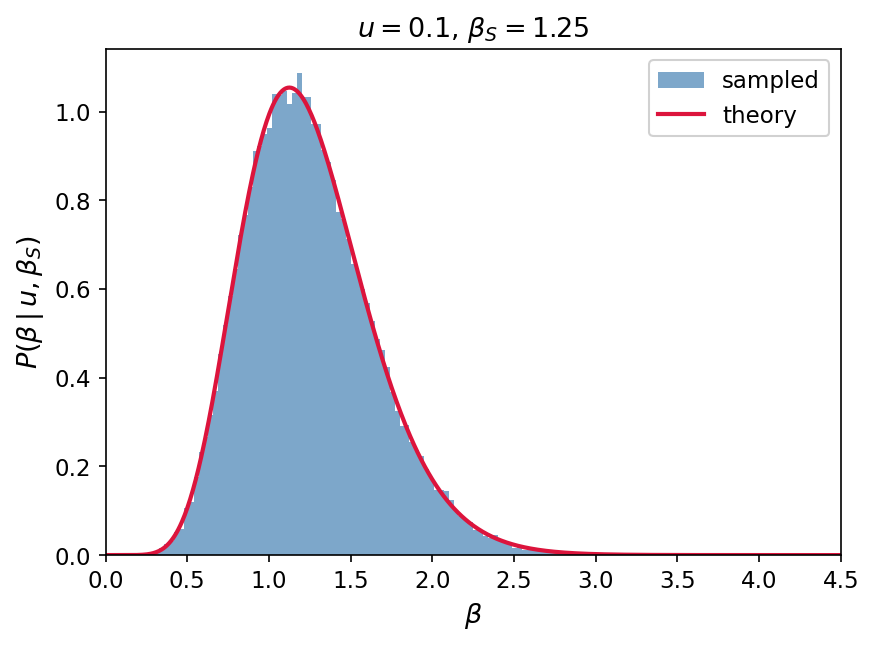}
\end{center}
\caption{Distribution of $\beta$ for $u=0.1$ and $\beta_S=1.25$.}
\label{fig:beta-histo-0p1}
\end{figure}

\begin{figure}[H]
\begin{center}
\includegraphics[width=0.6\textwidth]{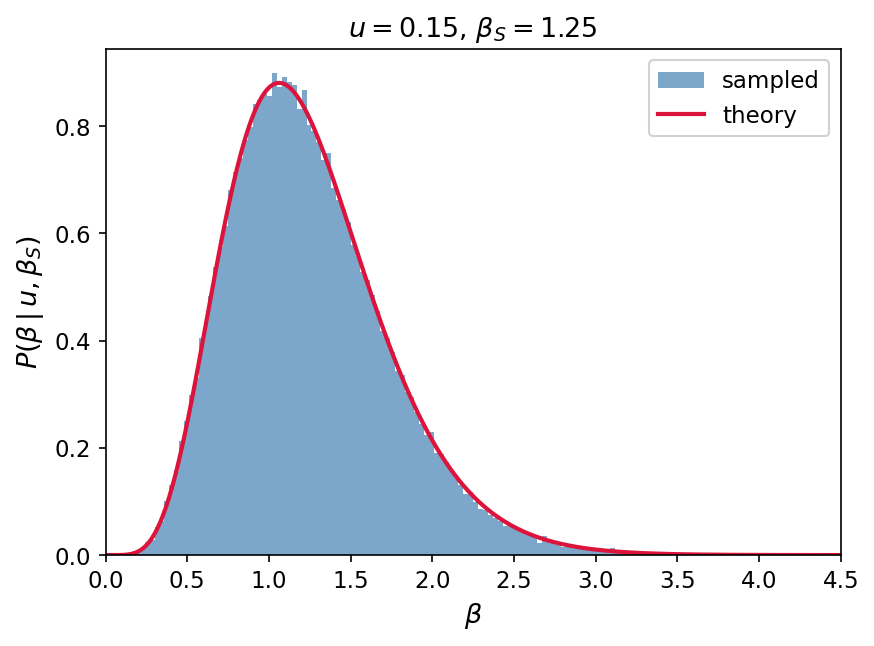}
\end{center}
\caption{Distribution of $\beta$ for $u=0.15$ and $\beta_S=1.25$.}
\label{fig:beta-histo-0p15}
\end{figure}

\begin{figure}[H]
\begin{center}
\includegraphics[width=0.6\textwidth]{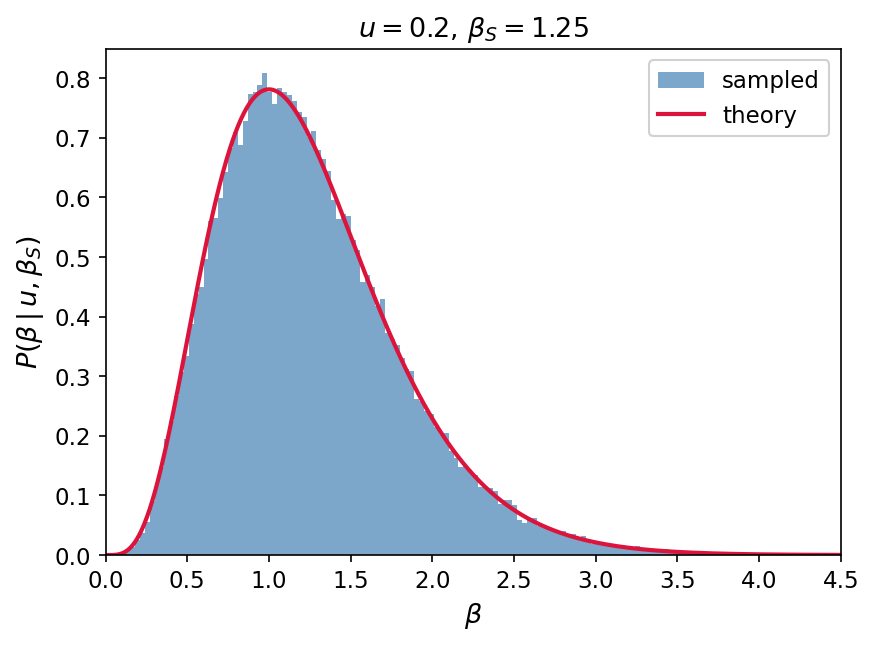}
\end{center}
\caption{Distribution of $\beta$ for $u=0.2$ and $\beta_S=1.25$.}
\label{fig:beta-histo-0p2}
\end{figure}

\begin{figure}[H]
\begin{center}
\includegraphics[width=0.6\textwidth]{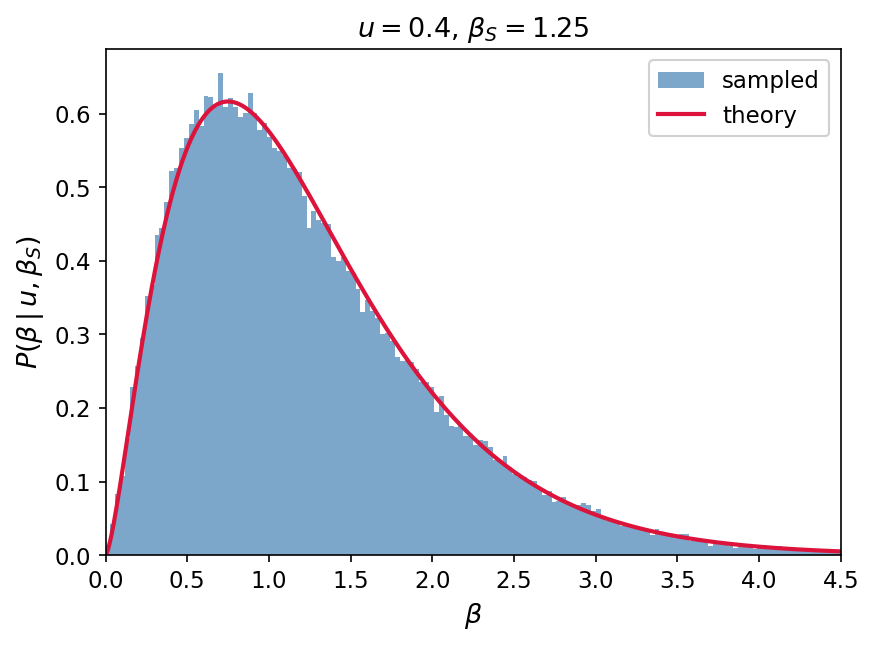}
\end{center}
\caption{Distribution of $\beta$ for $u=0.4$ and $\beta_S=1.25$.}
\label{fig:beta-histo-0p4}
\end{figure}

Although $\beta$ is not an observable, the values it takes along the chain can be compared with the superstatistical distribution $P(\beta|u,\beta_S)$ of Eq.~\eqref{eq:pbeta}, since this is the marginal distribution of $\beta$ in the joint distribution sampled by the chain. Figures~\ref{fig:beta-histo-0p01}--\ref{fig:beta-histo-0p4} show the histogram of the $\beta$ values collected along the chain ($N=10$, $10^5$ Gibbs steps after a warm-up of $10^3$), for the same values of $u$ and $\beta_S=1.25$.

The sample mean of $\beta$ reproduces $\beta_S$ within $0.6\%$ for every value shown, and its relative variance reproduces $u$ within $0.6\%$ for $u\le0.2$ and within $2.6\%$ at $u=0.4$, as expected from the definitions of $\beta_S$ and $u$ as the mean and the relative variance of $\beta$.

\subsection{Distribution of the single-particle kinetic energy}
\label{sec:moments-d}

\begin{figure}[H]
\begin{center}
\includegraphics[width=0.6\textwidth]{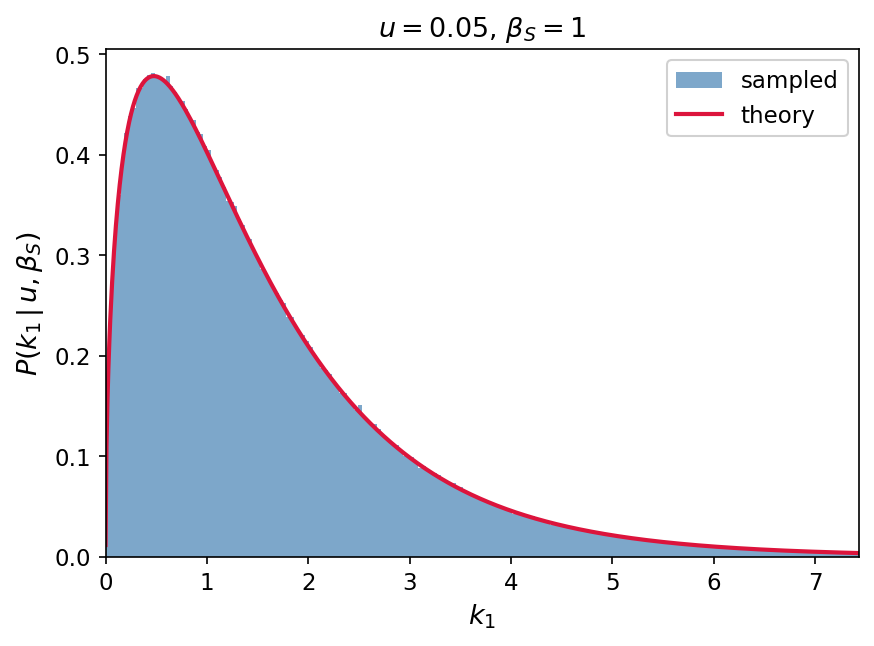}
\end{center}
\caption{Distribution of the single-particle kinetic energy for $u=0.05$.}
\label{fig:marginalpdf}
\end{figure}

Figure~\ref{fig:marginalpdf} compares the sampled kinetic energy of a single particle against the $\mathrm{BetaPrime}(3/2,1/u)$ density of Section~\ref{sec:obs-energy}, for $u=0.05$, $\beta_S=1$ and $M=10^6$ Gibbs steps. The sample mean kinetic energy is $1.5788$, against the theoretical value $1.5789$ of Eq.~\eqref{eq:mean-k-d}, that is, a relative error of $0.01\%$.

\subsection{Joint distribution of the kinetic energies of two particles}
\label{sec:res-joint}

\begin{figure}[H]
\begin{center}
\includegraphics[width=0.6\textwidth]{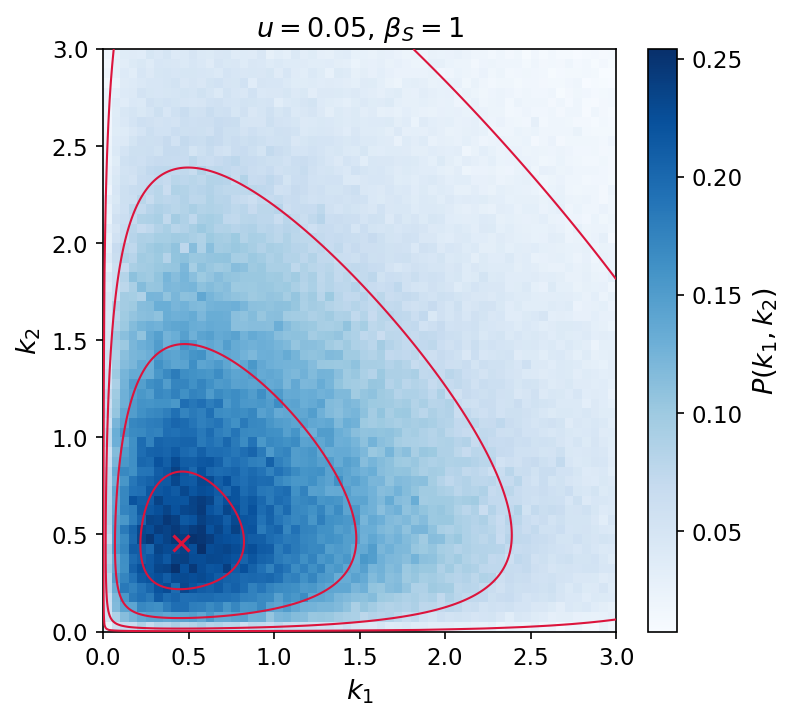}
\end{center}
\caption{Joint distribution of the kinetic energies $k_1$ and $k_2$.}
\label{fig:jointpdf}
\end{figure}

Figure~\ref{fig:jointpdf} shows the joint distribution of the kinetic energies of two particles, estimated from the pairs $(k_1,k_2)$ taken from the same sampled configuration $\bm V^{(t)}$ along the whole chain ($u=0.05$, $\beta_S=1$ and $M=10^6$ Gibbs steps), compared against Eq.~\eqref{eq:joint-3}. For these parameters the mode in Eq.~\eqref{eq:jointmode} is $k^*\approx0.45$, and the densest region of the histogram lies around this value, inside the innermost contour.

\subsection{Pearson correlation between the kinetic energies of two particles}
\label{sec:correlations}

\begin{figure}[H]
\begin{center}
\includegraphics[width=0.6\textwidth]{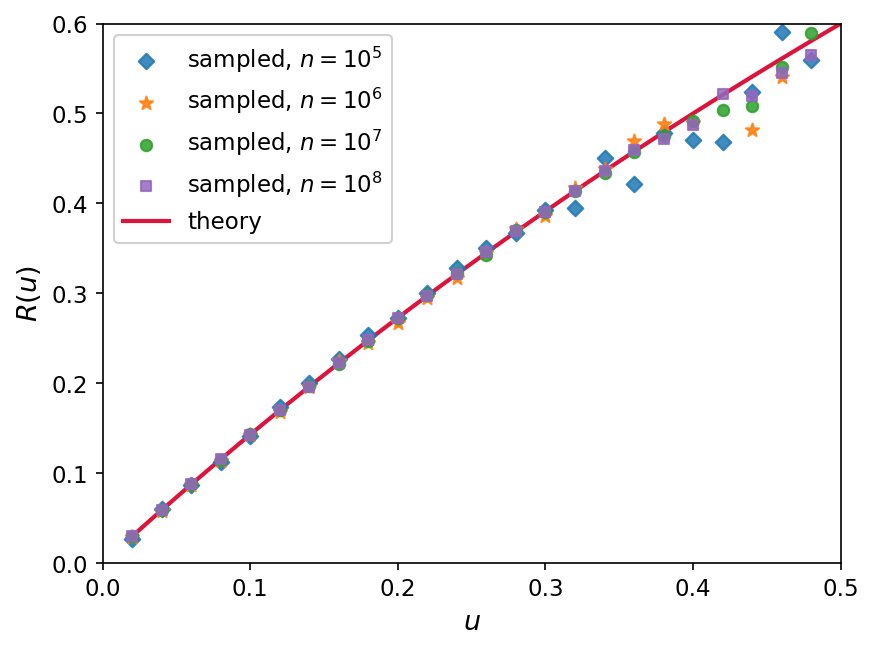}
\end{center}
\caption{Pearson correlation between $k_1$ and $k_2$ as a function of $u$.}
\label{fig:pearson}
\end{figure}

For $u=0.05$, $\beta_S=1$ and $M=10^6$, the sampler gives $\rho=0.07420$, against the theoretical value $0.07317$ (a relative error of $1.4\%$). Figure~\ref{fig:pearson} shows two regimes, separated by $u=1/4$. Below this value, the four sample sizes ($n=10^5$ to $10^8$) fall on the theoretical curve. Above it, the points scatter even for $n=10^8$, because the variance of the estimator $\hat R$ is finite only if $\langle k_1^2k_2^2\rangle<\infty$. By Eq.~\eqref{eq:mixed-moments} with $a=b=2$, this requires $u<1/4$. Above this value a single pair in the tail of the joint distribution can dominate the estimate; for instance, at $u=0.42$ and $n=10^6$ we obtain $\hat R=0.63$, against the theoretical value $0.52$.

\subsection{Mutual information}
\label{sec:mutinf}

\begin{figure}[H]
\begin{center}
\includegraphics[width=0.6\textwidth]{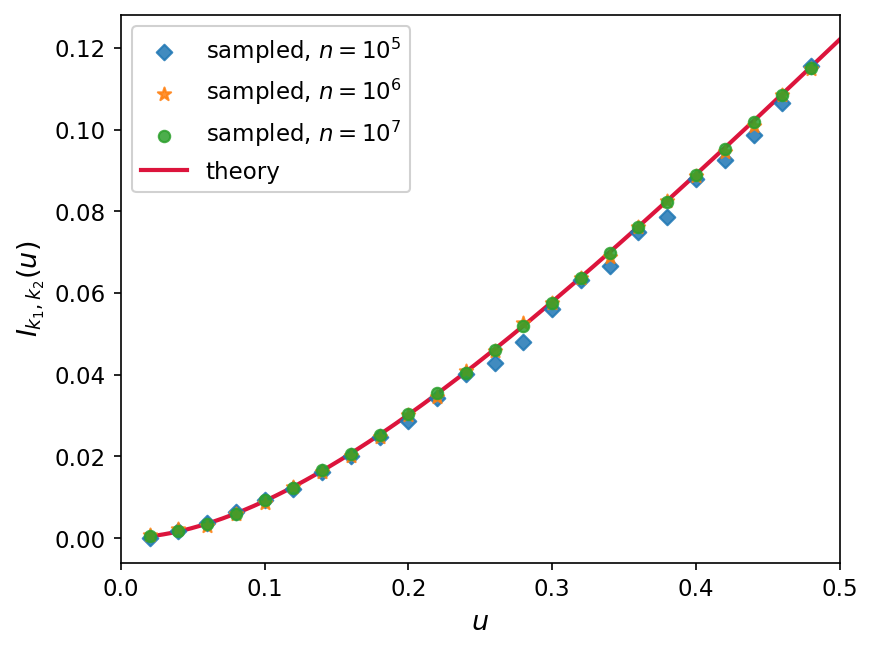}
\end{center}
\caption{Mutual information between $k_1$ and $k_2$ as a function of $u$.}
\label{fig:mutinf}
\end{figure}

Figure~\ref{fig:mutinf} shows the mutual information $I_{k_1,k_2}(u)$ measured on the sampled trajectories as a function of $u$, compared against Eq.~\eqref{eq:mutinf}. For $u=0.05$, $\beta_S=1$ and $M=10^6$, the $k$-nearest-neighbour estimate is $I_\text{est}=0.00222$~nats, against the theoretical value $0.00252$~nats of Eq.~\eqref{eq:mutinf}.

\section{Conclusions}
\label{sec:conclusions}

We have constructed a Gibbs sampler for the $N$-particle \emph{kappa} distribution within the framework of superstatistics. The sampler alternates between the conditional distribution of the velocities given $\beta$, a product of Maxwellians, and the conditional distribution of $\beta$ given the velocities, a gamma distribution. Both conditional distributions admit exact sampling, so the Gibbs sampler requires neither proposals nor rejection steps.

The $N$-particle \emph{kappa} distribution does not factorize into its single-particle marginals, because all the particles are described by a single distribution of $\beta$, integrated over once for the whole set. Hence single-particle quantities cannot distinguish it from independent particles, and its joint structure must be verified on two-particle quantities.

We verified numerically that the output of the sampler reproduces the analytical predictions at both levels: the \emph{kappa} distribution of velocities, the mean and relative variance of $\beta$, and the distribution of the single-particle kinetic energy, $u\beta_S k\sim\mathrm{BetaPrime}(3/2,1/u)$, with its mean. In particular, the expected correlations between the kinetic energies of two particles, described by their joint distribution, the Pearson correlation $R(u)=3u/(u+2)$ and the mutual information, are recovered in good agreement with the analytical expressions.

\section*{Acknowledgments}

This work was supported by ANID through Beca de Doctorado Nacional 2023 No.~21231822 (L.~H-F.) and FONDECYT grant 1220651 (S.~D.).

\section*{References}

\bibliography{gibbs}
\bibliographystyle{elsarticle-num}

\end{document}